\documentclass[11pt,a4paper]{article}

\usepackage{latexsym,amsmath,amssymb}
\usepackage{mathrsfs}
\allowdisplaybreaks
\renewcommand{\thefootnote}{\fnsymbol{footnote}}
\newcommand{\acknowledgments}{
$\\${\bf Acknowledgments}\newline}

\begin{document}
\begin{flushright}
\end{flushright}
\vspace{12mm}
\begin{center}
{{{\Large {\bf Note on two-dimensional gauged Lifshitz models}}}}\\[10mm]
{Edwin J. Son\footnote{email:eddy@sogang.ac.kr} and
Wontae Kim\footnote{email:wtkim@sogang.ac.kr}}\\[8mm]

{{Department of Physics, Sogang University, Seoul, 121-742, Korea\\[0pt]       
Center for Quantum Spacetime, Sogang University, Seoul, 121-742, Korea}\\[0pt]
}
\end{center}
\vspace{2mm}
\begin{abstract}
We fermionize the two-dimensional free Lifshitz scalar field 
in order to identify what the gauge covariant couplings are, and then 
they are bosonized back to get the gauged Lifshitz scalar field theories. 
We show that they give the same physical modes 
with those of the corresponding Lorentz invariant gauged scalar theories, 
although the dispersion relations are different.
\end{abstract}
\vspace{5mm}



\vspace{1.5cm}

\hspace{11.5cm}{Typeset Using \LaTeX}
\newpage
\renewcommand{\thefootnote}{\arabic{footnote}}
\setcounter{footnote}{0}

\section{Introduction}
A Lifshitz scalar field theory~\cite{lifshitz1,lifshitz2,hls,Horava:2008jf}
has been studied in the condensed matter physics as a description of
tricritical phenomena involving spatially modulated phases.  The
Lifshitz index $z$ reflects the anisotropic scaling between space and
time, $x\to bx$ and $t\to b^z t$, and the Lifshitz scalar theory
describes a free field fixed point $z=2$ in four dimensions.  
The Lorentz invariance emerges as an accidental symmetry at long
distances.  Recently, a renormalizable theory of gravity has been
also suggested by Ho\v{r}ava, so called Ho\v{r}ava-Lifshitz
gravity~\cite{horava,Horava:2009uw,Horava:2009if}.

On the other hand, as for the gauge coupling to the Lifshitz fields, 
it seems to be \textit{ad hoc} in that the gauge transformation of the {\it real} scalar field is arbitrary. 
Recently, we have obtained two-dimensional gauged Lifshitz scalar field theories by
considering the duality relation between the source current and the
Noether current~\cite{eks}. Now,  for a more systematic gauge coupling,
we want to take advantage of the two-dimensional theory, i.e., bosonization rules~\cite{coleman,mandelstam,halpern,Halpern:1975jc}.
First, the higher-derivative Lifshitz term can be lowered by introducing an auxiliary field, 
then all relevant terms become the first order derivative terms, 
which can be fermionized using the fermion-boson equivalence. 
However, we have to note that the first order form of the Lifshitz term 
produces a second class constraint~\cite{dirac}, but  one can make it 
the first class constraint system by adding counter-action without changing the physical contents.
Then, the advantage of the fermionic representation is that 
the minimal coupling to the gauge field is clear based on the gauge principle. 
In other words, we just change the ordinary derivative to the gauge covariant derivative 
for the minimal coupling. Once we get the gauged fermionic actions, 
then we recover the bosonic expression by means of the bosonization. 
In fact, there appear two types of couplings called the vector and the chiral,
where they exactly recover the bosonized Schwinger model for the vector case 
and the bosonized chiral Schwinger model for the chiral case for the vanishing Lifshitz term~\cite{twjz,bsg,jr,rajaraman,grr}.
In particular, for the latter case, 
the gauge anomaly appears from the chiral gauge coupling and the first order expression of the Lifshitz term simultaneously, 
so we have to determine the new Wess-Zumino action to cancel out the gauge anomaly~\cite{wz,fs,ht,fk,harada,kkk}.

In section~\ref{sec:sym}, the two-dimensional Lifshitz scalar term
can be formulated in terms of the first order form by introducing one auxiliary field; however, as the above mentioned,
we can maintain the first class constraint system by adding a counter-action to the first order Lifshitz action
without changing physical contents. 
In section~\ref{sec:LSM}, for the vector and the chiral couplings, 
it turns out that the final bosonic actions are just the vector and the chiral Schwinger models 
with the modified dispersion relations. In particular, for the chiral coupling, 
the bosonization process generates the gauge anomaly with one parameter gauge ambiguity. 
So, the Wess-Zumino action can be added to cancel out the gauge anomaly; however, 
it is slightly different from the conventional one because the symmetry breaking comes from the counter-action too. 
As a result, even for the chiral coupling, the gauged scalar theory can be consistently formulated.
Finally, conclusion and discussion will be given in section~\ref{sec:discussion}.
\section{First order form of Lifshitz scalar action}
\label{sec:sym}
The Lagrangian describing a Lifshitz real scalar field
$\phi$ up to the fourth derivatives is given as $
  \mathscr{L}_0(\phi) = \frac12 \left(\eta^{\mu \nu} \partial_\mu \phi \partial_\nu
    \phi + \beta \phi''^2 \right)$,
where $x^\mu = (t,x)$, and the metric is $\eta_{00} =
1$. The parameter $\beta$ is an arbitrary constant, which can be
restricted by based on physical requirements in later. The overdot and the
prime denote the derivatives with respect to $t$ and $x$,
respectively. Introducing an auxiliary field $\lambda$ for
conveniences, the Lagrangian density  can be written as~\cite{eks}
\begin{equation}
  \label{Lag:LS}
  \mathscr{L}_0(\phi, \lambda) = \frac12 \left[ \partial_\mu \phi \partial^\mu
    \phi + \beta (\lambda^2 - 2 \lambda' \phi') \right].
\end{equation}
Note that this is a  second class constraint system because the primary constraint 
$\Pi_\lambda \approx 0$ and the secondary constraint  $\beta(\lambda + \phi'') \approx 0$ 
does not commute. The simplest way to recover the broken local symmetry is to add
an action as $
  \mathscr{L}_\text{C}(\phi, \lambda, \Theta) = \frac12 \beta \left( 2 \lambda \Theta + \Theta^2 - 2 \Theta' \phi' \right)$~\cite{wz,fs,ht,fk}
Consequently, the total Lagrangian 
\begin{equation}
\begin{aligned}[b]
\mathscr{L}_\text{LB}(\phi,\lambda,\Theta) &= \mathscr{L}_0(\phi, \lambda) + \mathscr{L}_\text{C}(\phi,\lambda, \Theta) \\
  &= \frac12 \left[ \partial_\mu \phi \partial^\mu \phi + \beta (\lambda + \Theta)^2 - 2 \beta (\lambda' + 
\Theta') \phi') \right]
\end{aligned}
\end{equation}
 is invariant 
where the symmetry is implemented by $\delta \lambda = \xi$ and $\delta\Theta = -\xi$. It is natural to recover the original Lagrangian by choosing an appropriate gauge like the unitary gauge condition of
$\Theta \approx 0$, so that the reduced Hamiltonian becomes
$H_\text{red} = \frac12 \int dx \left( \Pi_\phi^2 + \phi'^2 + \beta \phi''^2 \right)$ where $\Pi_\phi$ is a conjugate momentum of $\phi$, and the Lifshitz parameter is assumed to be $\beta>0$ in order for the positivity of the reduced Hamiltonian.

The main goal is to obtain the gauged version of the Lifshitz scalar field theories;
however, it is not easy to achieve it in a systematic way. 
In particular, in two dimensions, there are
bosonization (or fermionization) rules~\cite{coleman,mandelstam,halpern,Halpern:1975jc}:
$\frac12 \partial_\mu \phi \partial^\mu
    \phi \leftrightarrow i \bar\Psi \gamma^\mu \partial_\mu \Psi, \quad 
    \epsilon^{\mu\nu} \partial_\nu \phi 
    \leftrightarrow \sqrt{\pi}
    \bar{\Psi} \gamma^\mu \Psi$,
especially, $\phi' \to \sqrt{\pi} \bar\Psi \gamma^0 \Psi$ with $\epsilon^{01}=+1$.
Then, the Lagrangian density of Lifshitz action and the counter-action can be written in terms of the fermionic variables as 
\begin{equation}
  \mathscr{L}_\text{LF}(\Psi, \lambda, \Theta) = i \bar\Psi \gamma^\mu \partial_\mu \Psi + \frac12 \beta \left( \lambda^2 - 2\sqrt{\pi} \lambda' \bar\Psi \gamma^0 \Psi \right) + \mathscr{L}_\text{C}(\Psi, \lambda, \Theta),
\end{equation}
where $\gamma^0=\sigma^1$, $\gamma^1=-i\sigma^2$, $\gamma^5=\gamma^0\gamma^1=\sigma^3$, and
$\mathscr{L}_\text{C}(\Psi, \lambda, \Theta) = \frac12 \beta \left( 2 \lambda \Theta + \Theta^2 - 2 \sqrt\pi \Theta' \bar\Psi \gamma^0 \Psi \right)$.

\section{Lifshitz-Schwinger Model}
\label{sec:LSM}

\subsection{Vector coupling}
\label{sec:GLS}

We note that it is easy to couple the gauge field to the fermionic variable with the help of the gauge symmetry. Now, considering the minimal coupling of the gauge field, the total Lagrangian density of gauged Lifshitz-fermion with the Maxwell term can be written as 
\begin{equation}
  \mathscr{L}_\text{VEC}= \mathscr{L}_\text{LF}(\Psi, \lambda, \Theta) + e A_\mu \bar\Psi \gamma^\mu \Psi
    - \frac14 F_{\mu\nu} F^{\mu\nu} \label{Lag:GLF:a}  
\end{equation}
where $D_\mu = \partial_\mu - i e A_{\mu}$ and $F_{\mu\nu} = \partial_\mu A_\nu - \partial_\nu A_\mu$. 
Then,  we can 
write the gauged bosonic version using the bosonization rules, which yields
\begin{equation}
  \label{Lag:GLS}
  \mathscr{L}_\text{VEC}= \mathscr{L}_\text{LB}(\phi, \lambda, \Theta) 
    + \frac{e}{\sqrt{\pi}} A_\mu \epsilon^{\mu\nu} \partial_\nu \phi - \frac14 F_{\mu\nu} F^{\mu\nu}.
\end{equation}
Note that the minimal coupling appears just at the fermionic kinetic term because the other couplings are already gauge invariant. 
From the Noether current, $
   J_\text{N}^\mu = \partial^\mu \phi - \beta \delta_1^\mu (\lambda'+\Theta') - \frac{e}{\sqrt\pi} \epsilon^{\mu\nu} A_\nu$,
which is not gauge invariant but conserved as $\partial_\mu J_\text{N}^\mu = 0$, one can define the gauge invariant current as
$  J^\mu \equiv J_\text{N}^\mu + \frac{e}{\sqrt\pi} \epsilon^{\mu\nu} A_\nu = \partial^\mu \phi - \beta \delta_1^\mu (\lambda'+\Theta')$,
which has an axial anomaly, $\partial_\mu J^\mu = (e/\sqrt\pi) \epsilon^{\mu\nu} \partial_\mu A_\nu$~\cite{twjz}

Next, using the equations of motion,
\begin{subequations}
\begin{align}
  & \Box \phi - \beta (\lambda''+\Theta'') - \frac{e}{\sqrt\pi} \epsilon^{\mu\nu} \partial_\mu A_\nu = 0, \\
  & \beta \left( \lambda + \Theta + \phi'' \right) = 0, \\
  & \partial_\mu F^{\mu\nu} - \frac{e}{\sqrt\pi} \epsilon^{\mu\nu} \partial_\mu \phi = 0,
\end{align}
\end{subequations}
we obtain one massive physical mode with the modified dispersion relation of
$  \left[ \Box + \beta \partial_x^4 + \frac{e^2}{\pi} \right] {}^*F = 0$
by eliminating the redundant variables of $\phi$, $\lambda$ and $\Theta$, where ${}^*F = \frac12 \epsilon^{\mu\nu} F_{\mu\nu} = -(e/\sqrt\pi) \phi$.
At last, the reduced Hamiltonian can be bounded $
  H_\text{red} = \frac12 \int dx \left( \Pi_\phi^2 + \phi'^2 + \beta \phi''^2 + \frac{e^2}
{\pi} \phi^2 \right)$, assuming $\beta>0$.

\subsection{Chiral coupling}
\label{sec:CLS}
Now, we are going to consider the chiral coupling, and the chiral gauged Lifshitz-fermion is given as
\begin{equation}
  \mathscr{L}_{\text{CHI}} =
    \mathscr{L}_\text{LF}(\Psi, \lambda, \Theta) + e A_\mu \frac{\left(1 - \gamma^5 \right)}{2} \bar\Psi \gamma^\mu \Psi 
    - \frac14 F_{\mu\nu} F^{\mu\nu}, 
\end{equation}
and its corresponding bosonic action becomes
\begin{equation}
  \label{Lag:CGLS}
  \mathscr{L}_{\text{CHI}} 
    = \mathscr{L}_\text{LB}(\phi,\lambda,\Theta) 
      + e  A_\mu \left( \eta^{\mu\nu} + \epsilon^{\mu\nu} \right) \partial_\nu \phi - \frac14 F_{\mu\nu} F^{\mu\nu} + \frac12 e^2 a A_\mu A^\mu
      + \mathscr{L}_\text{WZ},
\end{equation}
where we used the bosonization rule and changed the coupling $e \to 2e\sqrt\pi$ for convenience.
The mass term reflects the bosonization ambiguity $a$~\cite{jr}, and
the last term is new Wess-Zumino action to cancel the gauge anomaly 
\begin{equation}
  \label{Lag:WZ:new}
  \mathscr{L}_\text{WZ} = \frac12 (a-1) 
\partial_\mu \theta \partial^\mu \theta - e 
\partial_\mu \theta \left[ (a-1) \eta^{\mu\nu} + \epsilon^{\mu\nu} \right] 
A_\nu - \beta \theta' \left( \lambda' + \Theta' \right),
\end{equation}
where the gauge transformation is defined by $\delta A_\mu = (1/e) \partial_\mu \Lambda$, $\delta \phi = -\Lambda$, $\delta \theta = \Lambda$, and $\delta \lambda = \delta \Theta = 0$.
Note that the last term in Eq.~\eqref{Lag:WZ:new} comes from the Lifshitz term.
%
From the chiral action~\eqref{Lag:CGLS}, we obtain the equations of motion:
\begin{subequations}
\begin{align}
  & \Box \phi - \beta (\lambda''+\Theta'') + e \left( \eta^{\mu\nu} - \epsilon^{\mu\nu} \right) \partial_\mu A_\nu = 0, \\
  & \beta \left( \lambda + \Theta + \phi'' + \theta'' \right) = 0, \\
  & \partial_\mu F^{\mu\nu} + e \left( \eta^{\mu\nu} - \epsilon^{\mu\nu} \right) \partial_\mu \phi - e \left( (a-1) \eta^{\mu\nu} + \epsilon^{\mu\nu} \right) \partial_\mu \theta + e^2 a A^\nu = 0, \\
  & (a-1) \Box \theta - e \left( (a-1) \eta^{\mu\nu} + \epsilon^{\mu\nu} \right) \partial_\mu A_\nu - \beta \left( \lambda'' + \Theta'' \right) = 0.
\end{align}
\end{subequations}
After some calculations, they can be reduced to
$ \left[ \Box + e^2 a \right] \left[ (a-1) \Box + a \beta \partial_x^4 \right] \Phi 
  = -e^2 a \Box \Phi$
by eliminating ${}^*F$, $\theta$, $\lambda$ and $\Theta$, where $\Phi = \phi + \theta$. In fact, there are three primary constraints and the two secondary constraints with
the two gauge fixing conditions of $\theta \approx 0, \quad \Theta \approx 0$. The condition of $\theta =0$ requires one more stability condition, so that there exist ($6 \times 2$ variables)$-$(8 constraints)=($2 \times 2$ degrees of freedom) in the phase space. So, the physical dispersion relations in the Fourier modes are
$ k_0^2 
  = k_1^2 + \frac{a}{2(a-1)} \left[ (\beta k_1^4 + e^2 a) \pm \sqrt{[\beta k_1^4 - e^2 (a - 2)]^2 + 4 e^4 (a-1)} \right]$.
It shows that there are one massless mode from the ($-$) sign and one massive mode of $m^2 = e^2a^2/(a-1)$
from the ($+$) sign which is non-tachyonic for $a>1$. Actually, 
we defined the mass as a zero momentum limit of the energy. 
Now, it is natural to get the bounded reduced Hamiltonian,
\begin{equation}
\begin{aligned}
  H_\text{red} = \frac12 \int dx \bigg[ & \frac{1}{a-1} \left( \frac1e (\Pi^1)' + e A_1 + \Pi_\phi + \phi' \right)^2 + (\Pi^1)^2 \\
  & + \left( \Pi_\phi + e A_1 \right)^2 + \left( \phi' + e A_1 \right)^2 + \beta \phi''^2 + e^2 (a-1) A_1^2 \bigg],
\end{aligned}
\end{equation}
for $\beta>0$ and $a>1$.

\section{Discussion}
\label{sec:discussion}
We have studied how to obtain the gauge couplings in the Lifshitz scalar field theory in two dimensions.
Using the boson-fermion equivalence, the gauge coupling are easily realized.
It is interesting to note that the Lifshitz parameter $\beta$ should be positive for the positivity of the 
Hamiltonian. As a result, the gauged Lifshitz scalar theories possess the same physical modes with those of 
the conventional relativistic Schwinger and chiral Schwinger models, while the dispersion relations are different. 

As for the dispersion relation and the group velocity, 
there exist one massless mode and one massive mode for the chiral coupling case. 
The group velocity of the massless mode goes to the relativistic one $(\beta =0)$ asymptotically since
 $v_g^\text{massless} \approx 1 + O(k_1^{2})$ for the small $k_1$ and 
$v_g^\text{massless} \approx 1 + O(k_1^{-2})$ for the large $k_1$. In fact, It has one maximum and
one minimum group velocity so that it can be bounded. For the massive mode, as expected, 
the group velocity is monotonically increased like the Newtonian case. 
On the other hand, the energies ($\omega$) are always larger than those of the relativistic ones because of the positive 
contribution to the dispersion relation.  

Finally, 
one might wonder how the massive modes appear and what the underlying mechanism for the mass 
generation is in these two models? Note that the masses are
 in fact independent of the Lifshitz coupling $\beta$. Moreover, the gauged vector and chiral Lifshitz models 
clearly become the bosonized Schwinger model and the bosonized chiral Schwinger model for $\beta=0$,
 respectively. It means that the mass generation has something to do with the mass generation 
of the original models without Lifshitz terms.  For the vector coupling, the vector current should 
be written in terms of the gauge invariant fashion to keep the gauge invariance in the bosonized version 
which corresponds to the quantized version of the original fermionic theory. Thanks to the (local) gauge invariance
 and the conservation of the vector current, the gauge invariant axial current becomes anomalous. 
In this case, the global symmetry has been broken. By the way, as for the case of the chiral coupling, 
the theory becomes anomalous after quantization, i.e., the chiral current which is a source current is not 
conserved because of lack  of the local gauge invariance. It is a crucial difference between the vector 
and the chiral models. Thus, different types of anomaly called the axial anomaly for the former case
and the chiral anomaly for the latter case are related to each mass generation, which consequently 
affect the equations of motion for the gauge field. So, the mass appears in the gauge-invariant fashion
for the vector case while it appears in the gauge-noninvariant fashion for the chiral case.
In particular, one can note that the mass parameter is connected 
with the coupling constant because the coupling $e$ has the mass dimension only in two dimensions.
Unfortunately, it is not easy to realize the above anomaly argument for the mass generation 
up to four dimensional theories even in spite of the similarity of anomalous structures. 

\acknowledgments
E.\ J.\ Son was supported by the National Research Foundation of
Korea(NRF) grant funded by the Korea government(MEST) through the
Center for Quantum Spacetime(CQUeST) of Sogang University with grant
number 2005-0049409.
W.\ Kim was supported by the National Research Foundation of Korea(NRF) grant funded 
by the Korea government (MEST) (2011-0005147).


\end{document}